\documentclass[epj]{svjour}
\usepackage{graphics}
\begin{document}

\title{Editorial: New Frontiers in Holographic Duality}
\subtitle{From quantum complexity and black holes to hydrodynamics and neutron stars}
\author{Ayan Mukhopadhyay  
}                     
\offprints{}          
\institute{Center for Strings, Gravitation and Cosmology, Indian Institute of Technology Madras, Chennai 600 036, India}
\date{}
%
\abstract{
Over the last twenty five years, holographic duality has revolutionised our understanding of gauge theories, quantum many-body systems and also quantum black holes. This topical issue is a collection of review articles on recent advances in fundamentals of holographic duality and its applications with special focus on a few areas where it is inter-disciplinary to a large measure. The aim is to provide a sufficient background on relevant phenomenology and other theoretical areas such as quantum information theory to researchers whose primary expertise is in quantum fields, strings and gravity, and also the necessary concepts and methods of holography to researchers in other fields, so that these recent developments could be grasped and hopefully further developed by a wider community. The topics relating to fundamental aspects include understanding of bulk spacetime reconstruction in holography in the framework of quantum error correction along with the spectacular advances in resolution of the information paradoxes of quantum black holes; quantum complexity and its fundamental role in connecting holography with quantum information theory; theoretical and experimental advances in quantum simulators for information mirroring and scrambling in quantum black holes, and teleportation via wormholes; and a pedagogical review on wormholes also. The topics related to applied holography include applications to hydrodynamic attractor and its phenomenological implications, modelling of equation of state of QCD matter in neutron stars, and finally estimating hadronic contribution to light-by-light scattering for theoretical computation of the muon's $g-2$.
} 
\maketitle
\section{Introduction}
\label{intro}
ADS/CFT a.k.a. gauge/gravity duality \cite{Maldacena:1997re} has proved to be a landmark in theoretical physics. It is the first concrete realization of the holographic principle of quantum gravity according to which the dynamics of quantum spacetime can be described in terms of degrees of freedom living at the boundary. The duality provides precise examples in which quantum gauge theories in $d$-dimensions are dual to string theories in backgrounds such as $AdS_{d+1}\times X$ where $X$ is a suitable compact space. Remarkably, this is a weak/strong duality. When the gauge theory is strongly coupled and the gauge group has a large rank, the dual string theory reduces to semiclassical Einstein's gravity in one higher dimension coupled to a few fields. On the other hand, perturbative processes in gauge theories should yield understanding of quantum stringy spacetime. The mysteries of strongly coupled dynamics such as confinement, hydrodynamization far away from equilibrium, phases of matter without quasiparticles, and solving the dynamics in real time can be mapped to appropriate questions in classical gravity. The duality promises resolutions of the information paradoxes of black holes, one of the most longstanding mysteries of theoretical physics, via better understanding of how quantum spacetime is encoded into the dynamics of the degrees of freedom of the dual gauge theories.

Over the last twenty five years, there has been remarkable progress in applications of this duality to deeper understanding of quantum field theories \cite{Aharony:1999ti} with applications to Quantum Chromodynamics (QCD) \cite{Kim:2012ey,Johanna} and also strongly correlated quantum matter \cite{Hartnoll:2016apf}. Such developments have left a very strong influence over research in fields as diverse as particle phenomenology, condensed matter physics, heavy-ion collisions, and more recently in QCD matter in neutron stars and also cosmology. Furthermore, these applications have opened up new fundamental questions into the nature of quantum matter, such as how the limits of applicability of hydrodynamics get pushed to even small systems and systems that are far from equilibrium \cite{Hubeny:2010ry,Heller:2016gbp}, relationship between transport and quantum scrambling with applications to the understanding of strange metals \cite{Jahnke:2018off,Samanta:2022rkx}, existence of new mechanisms for superconductivity which do not require quasiparticles \cite{Hartnoll:2008kx}, existence of non-Fermi liquid fixed points for fermions at finite density in higher dimension \cite{Iqbal:2011ae} and also the dynamics of quantum entanglement in field theories \cite{Liu:2013iza}. Since almost all such questions can be mapped to the nature of near-horizon geometries of quantum and classical black holes, this duality has turned out to be one of the most interdisciplinary areas in theoretical physics.

More recently, the introduction of tools of quantum information theory has revolutionized our understanding of this duality. The fundamental inputs here have been precise proposals of how extremal surfaces in the emergent spacetime capture entanglement (or more precisely quantum entropic measures) in the dual field theory \cite{Ryu:2006bv,Hubeny:2007xt,Engelhardt:2014gca}. The encoding of subregions of the emergent spacetime into the degrees of freedom of the dual field theory has now been precisely formulated in terms of a quantum error correcting map especially in the limit where gravity admits a semi-classical description \cite{Harlow:2014yka}. Constructions of novel quantum circuits \cite{Pastawski:2015qua} with such error correcting properties have elucidated how the bulk spacetime region bounded by the extremal surface (a.k.a. entanglement wedge) can be encoded into the boundary degrees of freedom without any contradiction with properties of von Neumann algebras. 

In the last three years, these developments have been applied to achieve a major breakthrough in our understanding of quantum black holes, particularly it has been shown how the duality can be used to reproduce the Page curve of the Hawking radiation from the semiclassical evaporating black hole geometry \cite{Almheiri:2020cfm}. Although we do not yet fully understand how principles like black hole complementarity can emerge from the microscopic description (or equivalently dynamics of the black hole microstates), we better understand how to compute fin(er) grained entropy of the Hawking radiation of an evaporating black hole in a way which is compatible with unitarity -- a major step forward in resolving the original information loss paradox by Hawking. Perhaps even more exciting is how such developments have led to deeper understanding of the fundamental role of quantum information in quantum gravity. In fact we can now even hope that quantum circuits can simulate many aspects of dynamics of quantum spacetime, and lead to new insights. 

The applications of gauge/gravity duality have grown in the meanwhile covering diverse fields where the role of phenomenology is crucial. More excitingly, these lead to the possibility of posing new fundamental questions on quantum matter, hadronic structure, etc and newer arenas where concrete confrontations with experiments are possible.

The present volume of topical reviews is motivated by the need of bringing diverse communities together particularly in a way that allows understanding of multiple perspectives and modes of thoughts. It is clear that the potential of researchers in diverse fields such as quantum information and condensed matter contributing to the understanding of deep aspects of quantum gravity leading to major new breakthroughs is materializing rapidly. Similarly, the potential for gauge/gravity duality making fundamental impact into understanding of quantum matter with successful confrontations with experiments is expanding. However the production of reviews which can lead to more rapid mutual understanding of these perspectives based on which the interdisciplinary questions are framed and investigated is not keeping pace with the developments in fundamentals and applications of gauge/gravity duality. The review articles in this collection attempt to address this by focusing in some key directions. 

The reviews appearing in this volume are loosely divided into two categories. The reviews in the first category address the developments at the intersection of quantum information and gravity including the recent breakthroughs related to progress in resolution of black hole paradoxes and the understanding of spacetime reconstruction from dual field theory. The reviews in the second category addresses major developments in applied gauge/ gravity duality which are of topical/current interest and where understanding of both phenomenological methods and the tools of gravity are crucial. We briefly describe the content of these reviews in what follows.

\section{Quantum information and gravity}
This volume has four reviews on topics at the intersection of quantum information and gravity. The review of Chapman and Policastro \cite{Chapman:2021jbh} is on the newest entry into the holographic dictionary, quantum complexity. A precise formulation of the holographic dictionary for quantum complexity is still under progress and is a very interdisciplinary topic of research. Nevertheless, there has been a lot of developments which have already had an impact of our understanding of quantum information processing of black holes, especially in relation to the role of complexity in resolving the Almheiri-Marolf-Polchinski-Sully (AMPS) paradox which questions the validity of the black hole complementarity paradigm. This pedagogical review introduces the relevant backgrounds in quantum information theory and gravity, critically examines the successes of current proposals, and discusses the open questions.

The review by Kibe, myself and Mandayam \cite{Kibe:2021gtw} delves into the recent progress into holographic spacetime reconstruction in the framework of quantum error correction, and the related recent breakthroughs in understanding of resolution of information paradoxes of black holes. A critical discussion on microstate models and their role in elucidating how black hole complementarity principle can emerge from microscopics without encountering any paradoxes is a very important part of this review. The review also has a pedagogical introduction to quantum error correction and also aspects of gravity which should be useful for beginning researchers. 

The review by Bhattacharyya, Joshi and Sundar \cite{Bhattacharyya:2021ypq} focuses on the recent progress in quantum simulations of various aspects of quantum black holes such as quantum information scrambling and information mirroring by black holes past their Page time, and also the teleportation protocols realized by traversable wormholes. This examines possibilities of learning new aspects of quantum gravity from such quantum simulators. There is equal focus on the interdisciplinary concepts, and also describes details of some actual realizations of such quantum simulators in the laboratory.

The review by Kundu \cite{Kundu:2021nwp} focuses on wormholes which plays ubiquitous and yet diverse roles at the intersection of quantum information and quantum gravity. The review complements the two reviews on spacetime reconstruction and black holes, and quantum simulations of spacetime dynamics by elaborating on the gravitational constructions of wormholes in a pedagogical way. It also discusses some of the puzzles that wormholes create in the fundamental formulation of the gauge/gravity duality itself, which is one of the most active topic of discussion in this field. Beginning researchers should find this review useful as it provides a much needed guide to understand constructions of wormholes for diverse applications.

\section{Applied holography}
This volume contains three reviews in this category. The review by J\"arvinen \cite{Jarvinen:2021jbd} focuses on the applications of gauge/gravity duality in understanding the nature of dense and cold QCD matter in neutron stars. It is a state of the art account of how diverse tools like holography, effective field theory, lattice QCD and perturbative QCD can be combined together with phenomenological inputs from gravitational wave astronomy to unravel one of the most inaccessible forms of quantum matter whose understanding poses a formidable challenge in theoretical physics. 

The review by Leutgeb, Mager and Rebhan \cite{Leutgeb:2021bpo} discusses the role of holography in understanding the hadronic contribution of the light-by-light scattering which plays a fundamental role in recent theoretical efforts for the calculation of the anomalous magnetic moment of the muon. This topic is of utmost importance in understanding whether the experimental results for the anomalous magnetic moment of the muon can be explained within the Standard Model of particle physics.

Finally the review by Soloviev \cite{Soloviev:2021lhs} focuses on one of the most important discoveries in physics of out-of-equilibrium systems pioneered by holography, namely hydrodynamic attractors which explain why and how hydrodynamics can apply to systems far away from equilibrium with large pressure anisotropies. The influence of these discoveries on perturbative approaches has been discussed along with phenomenological applications to heavy-ion collisions also. The review also discusses new hybrid approaches \cite{Mitra:2020mei} which could be essential in the latter context.

\section{Hopes}

We hope that the present volume would be useful for researchers in diverse fields who want to contribute to the understanding of quantum gravity, and also for \textit{holographers} (experts on the tools of gauge/gravity duality) to create new impact into understanding of extreme and novel quantum matter, especially with confidence to confront experiments. We particularly hope that the articles will be useful for beginning researchers both in terms of pedagogical introductions into different conceptual paradigms, and theoretical/phenomenological tools, and also for getting a first understanding of diverse interlinked perspectives and cultures within theoretical physics that could be crucial for interdisciplinary research necessary for some fundamental discoveries.

%
%
%
%

\end{document}